# Superconductivity in NdFe$_{1-x}$Co$_x$AsO (0.05 < x < 0.20) and rare-earth magnetic ordering in NdCoAsO

Andrea Marcinkova[1], David A. M. Grist[1], Irene Margiolaki[2], Thomas C. Hansen[3], Serena Margadonna[1] and Jan-Willem G. Bos[1,4]

1. School of Chemistry and Centre for Science at Extreme Conditions, University of Edinburgh, United Kingdom, EH9 3JJ.

2. European Synchrotron Radiation Facility, 39043 Grenoble, France.

3. Institute Laue Langevin, 38042 Grenoble, France.

4. Department of Chemistry – EPS, Heriot-Watt University, Edinburgh, United Kingdom, EH14 4AS

The phase diagram of NdFe$_{1-x}$Co$_x$AsO for low cobalt substitution consists of a superconducting dome (0.05 < x < 0.20) with a maximum critical temperature of 16.5(2) K for x = 0.12. The x = 1 end member, NdCoAsO, is an itinerant ferromagnet (T$_C$ = 85 K) with an ordered moment of 0.30(1) µ$_B$ at 15 K. Below T$_N$ = 9 K, Nd spin-ordering results in the antiferromagnetic coupling of the existing ferromagnetic planes. Rietveld analysis reveals that the electronically important two-fold tetrahedral angle increases from 111.4º to 115.9º in this series. Underdoped samples with x = 0.046(2) and x = 0.065(2) show distortions to the orthorhombic Cmma structure at 72(2) and 64(2) K, respectively. The temperature dependences of the critical fields H$_{c2}$(T) near T$_c$ are linear with almost identical slopes of 2.3(1) T K$^{-1}$ for x = 0.065(2), x = 0.118(2) and x = 0.172(2). The estimated critical field H$_{c2}$(0) and correlation length for optimally doped samples are 26(1) T and 36(1) Å. A comparison of the maximum reported critical temperatures of well-characterized cobalt doped 122- and 1111-type superconductors is presented.

## Introduction

The 2008 discovery[1] of high-$T_c$ superconductivity based on iron arsenide layers has generated enormous interest (Refs 2, 3 and references therein). Presently, the most widely investigated systems are the 1111-type RFeAsO (R = La, rare-earth) and 122-type AEFe$_2$As$_2$ (Ae = alkaline earth) superconductors. Both contain square planar Fe layers, with Fe tetrahedrally coordinated by As, which are kept apart by either RO layers or AE ions. The 1111-type superconductors were the first to be discovered and hold the current record for critical temperatures and fields (up to 55 K and of the order of 100 Tesla), while single crystals of the 122's ($T_c$ up to 38 K) are readily available making detailed physical property studies possible. Like the cuprates, the iron arsenides have an antiferromagnetic (AF) parent. In contrast to the cuprates, the parent materials are metallic and not Mott insulators. The magnetism is therefore not due to superexchange but to Fermi surface nesting in the 2-dimensional bandstructure.[4, 5] The parent materials are rendered superconducting (SC) via chemical doping[2, 3] or in some cases application of hydrostatic pressure.[6] Both electron and hole doped superconductors exist.[2, 3] Most temperature-composition phase diagrams show a dome-shaped SC region, and suggest the coexistence of magnetic order and superconductivity for underdoped samples.[7-11] An exception is the phase diagram of LaFeAsO$_{1-x}$F$_x$ where a discontinuous transition into the superconducting state is reported.[12] In the cuprate high-$T_c$ superconductors, chemical substitutions on the copper site are detrimental to superconductivity. By contrast in the iron arsenides, direct substitution on the Fe site with a variety of transition metals is known to induce superconductivity. This was first reported for 1111-type LaFe$_{1-x}$Co$_x$AsO with a maximum critical temperature of 13-14 K.[13] Other well characterized cobalt doped superconductors include, 122-type AE(Fe$_{1-x}$Co$_x$)$_2$As$_2$ (AE = Ba, Sr and Ca)[14-16], 1111-type SmFe$_{1-x}$Co$_x$AsO,[17] and 1111-type CaFe$_{1-x}$Co$_x$AsF.[18] The highest reported $T_c$ for cobalt doping is 22 K, and is found for Ba(Fe$_{1-x}$Co$_x$)$_2$As$_2$ and CaFe$_{1-x}$Co$_x$AsF. The two-fold tetrahedral As-Fe-As angle ($\alpha$) is an important electronic parameter with higher $T_c$'s found as $\alpha$ tends to the ideal cubic value (109.5°).[8, 19] Most forms of chemical doping result in a reduction of α and values closer to 109.5°. For example, indirect electron doping of NdFeAsO (α = 111.4°) via F substitution or O deficiency reduces α by ~ 0.4° and ~ 1° for optimally doped samples.[19, 20] This is also true for hole doped 122 systems such as $Ba_{1-x}K_xFe_2As_2$ where α = 109.5° for samples with $T_c$ = 38 K (x = 0.4), compared to α = 111.2° for the parent material.[21] In contrast, our results show that α increases with cobalt substitution (~1° for x = 0.12), which is expected to have an unfavorable effect and reduce $T_c$ from what is possible for a given nominal doping level (x). This effect is common to all cobalt doped superconductors but does not seem to have been explicitly pointed out in the literature.

The fully substituted, x = 1, samples are also of considerable interest. LaCoAsO is an itinerant ferromagnet[13, 22] with an anomalous magnetization similar to MnSi and $Fe_xCo_{1-x}Si$.[23] $BaCo_2As_2$ is not ferromagnetic (FM) but is in close proximity to a FM quantum critical point.[24] The presence of Nd in the current x = 1 composition offers the prospect of studying the interplay between rare-earth and transition metal magnetism. In case of the RFeAsO parent materials this has generated much interest and a variety of combined R and Fe magnetic structures have been reported.[25, 26] In all cases, $T_N$ of the R sublattice is much lower than $T_{SDW}$ (~ 140 K) for the Fe magnetic order, and the weak R-Fe interaction is not considered crucial in explaining the higher $T_c$'s observed for magnetic R ions.[20, 25]

In this manuscript, the temperature-composition phase diagram for $NdFe_{1-x}Co_xAsO$ (0 ≤ x ≤ 1) is reported from a combination of synchrotron X-ray, neutron powder diffraction, magnetic susceptibility and electrical resistivity measurements. These measurements reveal a SC dome extending from 0.05 < x < 0.20 with maximum $T_c$ = 16.5(2) K and $H_{c2}(0)$ = 26(1) T for x = 0.12. NdCoAsO, is an itinerant ferromagnet ($T_C$ = 85 K) that shows a transition to long range AF order ($T_N$ = 9 K) upon ordering of the rare-earth spins. This work follows on from our earlier investigations into the effects of electron (via F substitution)[20, 27] and hole doping (via Ca and Sr substitution)[28, 29] of the parent material NdFeAsO.

**Experimental**

Polycrystalline samples of the NdFe$_{1-x}$Co$_x$AsO ($x_{nominal}$ = 0, 0.05, 0.075, 0.10, 0.125, 0.15, 0.175, 0.20, 0.25, 0.50 and 1) were prepared using standard solid state chemistry methods. Stoichiometric mixtures of NdAs, Fe$_2$O$_3$, Fe, Co$_3$O$_4$ and Co powders of at least 99.9% purity were mixed using mortar and pestle and pressed into pellets. The pellets were vacuum sealed into quartz tubes and heated for 48 hours at 1150 °C for 0 ≤ $x_{nominal}$ ≤ 0.25 and at 1050 °C for $x_{nominal}$ = 0.5 and $x_{nominal}$ = 1. The reactions were initially done on a 0.5 gram scale. Larger 2 gram samples of $x_{nominal}$ = 0.075, 0.125, 0.175 and 1 were prepared for neutron powder diffraction. The starting material NdAs was prepared by heating stoichiometric mixtures of Nd and As pieces at 850 °C for 2 x 12 hours with an intermediate shaking of the sealed tube. Initial phase analysis was done using laboratory powder X-ray diffraction on a Bruker D8 AXS diffractometer with a Cu K$_{\alpha 1}$ radiation source. Zero field cooled (ZFC) and Field Cooled (FC) DC magnetic susceptibilities were measured using a Quantum Design Magnetic Property Measurement System (MPMS). The applied magnetic field (H) for the superconductivity tests was 20 Oe. The temperature dependence of the susceptibility of the $x_{nominal}$ = 1 sample was measured in H = 10 kOe. The temperature and field dependences of the electrical resistivity were measured using the resistance option of a Quantum Design Physical Property Measurement System (PPMS). The resistivities were measured using the four point contact method on bars of approximately 1 x 1 x 5 mm$^3$. Sintered samples tend to degrade into powder when not kept in a closed container. This does not affect the crystallinity or the diamagnetic response but prevented us from obtaining resistivity measurements for some of the reported samples. High-resolution synchrotron powder diffraction measurements were done on the ID31 beamline at the European Synchrotron Radiation Facility in Grenoble. The X-ray wavelength used was 0.3994 Å, and data were binned with a 0.02° stepsize between 0 ≤ 2θ ≤ 35°. The samples were contained in 0.5 mm diameter silica capillaries. Room temperature patterns were collected for all prepared samples while selected samples ($x_{nominal}$ = 0.05, 0.075, 0.125, 1) were studied as a function of temperature using shorter scans. Room temperature neutron powder diffraction patterns for $x_{nominal}$ = 0.075, 0.125, 0.175 and 1 were collected on the D20 beamline[30] at the Institute Laue Langevin in Grenoble. The instrument was used in the high-flux setting with and $\lambda$ = 1.304 Å (42° take-off angle, Cu monochromator). The samples were contained in 6 mm diameter vanadium cans and cooled in a standard orange cryostat. Variable temperature data were collected between 1.7 K and 100 K for NdCoAsO at wavelength $\lambda$ = 2.419 Å (42° take-off angle, pyrolytic graphite monochromator). Data were collected between $10 \leq 2\theta \leq 140°$ and binned with a 0.1° stepsize. Rietveld analysis of the collected powder diffraction data was done using the GSAS suite of programs.[31] A pseudo-Voigt function using Stephens anisotropic peak broadening was used to describe the peak shape for the synchrotron X-ray diffraction data.[32]

**Structure and properties of NdCoAsO (x = 1).**

The temperature dependence of the crystal structure of NdCoAsO was followed between room temperature (RT) and 5 K using synchrotron X-ray powder diffraction. No structural transitions were observed, and the structure was described using the P4/nmm structural model reported by Quebe et al.[33] The RT lattice constants, atomic parameters, selected bond lengths and fit statistics are summarized in Table 1. The refined lattice constants and fractional coordinates at 5 K are: $a$ = 3.97940(1) Å, $c$ = 8.29849(3) Å, $z_{Nd}$ = 0.14264(4) and $z_{As}$ = 0.65071(7) [$\chi^2$ = 4.1, $wR_p$ = 9.5 %, $R_p$ = 6.2 %, $R_F^2$ = 2.4%]. The Co-As bond distance contracts moderately from 2.3546(3) Å at RT to 2.3501(3) Å at 5 K, while the tetrahedral angle remains almost constant with RT and 5 K values of 115.85(6)° and 115.69(4)°, respectively.

The DC magnetic susceptibility is shown in Fig. 1a and reveals a FM divergence of the susceptibility with $T_c$ = 85 K, followed by a transition to an AF state at $T_N$ = 9 K. The Curie temperature was determined from the local maximum in $d\chi/dT$, while the Neel temperature was taken when $d\chi/dT$ changes sign (Fig 1a). Zero field cooled and field cooled curves collected in 1 Tesla do not show any thermal hysteresis. The high-temperature susceptibility does not follow the Curie-Weiss law. To further investigate the magnetic properties, M(H) isotherms were collected at 1.7 K, 50 K and 100 K. These data are shown in the inset of Fig. 1b. The main panel shows the derived Arrott plots, $M^2$ versus H/M, that were used to determine the nature of the magnetic ground state.[34] Extrapolation of the linear high-field behavior for a ferromagnet yields a negative H/M-axis intersect, while the same extrapolation for AF and paramagnetic (PM) states intersect the H/M-axis at a positive value. At 1.7 K and 100 K, the Arrott plots clearly indicate the absence of ferromagnetism, and are consistent with antiferromagnetism and paramagnetism, respectively. At 50 K, the Arrott plot indicates a FM ground state. The isothermal M(H) measurements are therefore consistent with a PM to FM to AF sequence of ordering transitions upon cooling from RT. No magnetic hysteresis is evident from the measurement at 50 K indicating that NdCoAsO is an extremely soft ferromagnet without any sizable remnant magnetization.

The temperature dependence of the electrical resistivity of NdCoAsO is typical of that of a good metal (RRR = 25, Fig. 1a). The magnetic ordering transitions do not result in large anomalies in R(T) but are evident in dR/dT, which shows a maximum at ~85 K and a minimum at ~ 9 K (Fig. 1a). The slope remains positive throughout revealing the sample is metallic over the entire temperature range.

The magnetic ordering in NdCoAsO was further investigated using high-flux neutron powder diffraction on a 2 gram polycrystalline sample. Patterns were collected for two hours each at 1.7 K, 15 K, 40 K, 70 K and 100 K. At 1.7 K, long range AF order was confirmed by the presence of magnetic Bragg reflections that were not present in the 5 K synchrotron powder diffraction pattern. These reflections were all indexed on a tetragonal cell doubled along the crystallographic c-direction ($a_m = a_N$, $c_m = 2c_N$, where the subscript m denotes magnetic and N the nuclear cell). The indexing of the most prominent magnetic reflections is given in Fig. 1c. Upon heating to 15 K, the AF reflections disappear, in agreement with the magnetic susceptibility data. Careful subtraction of the 100 K and 15 K data revealed weak magnetic contributions to the nuclear (001) and (002) Bragg reflections. Similar FM intensities were also evident in the 40 K pattern but at 70

K there is no evidence for magnetic Bragg diffraction. This apparent disagreement with susceptibility data is most likely due to the small ordered FM cobalt moment close to $T_C$. The possible FM and AF magnetic structures were analyzed using representational analysis. These calculations were performed using version 2K of the program SARAh representational analysis.[35] In both the FM ($\mathbf{k} = 0$) and AF ($\mathbf{k} = (0\ 0\ ½)$) states all 16 symmetry elements of the P4/nmm space group leave the magnetic propagation vector ($\mathbf{k}$) invariant or transform it into an equivalent vector, and thus constitute the small group $G_\mathbf{k}$. In the FM state, the decomposition of the magnetic representation $\Gamma_{Mag}$ in terms of the irreducible representations (IRs) of $G_\mathbf{k}$ for the Co site is $1\Gamma_3^1 + 1\Gamma_6^1 + 1\Gamma_9^2 + 1\Gamma_{10}^2$. The representations used are after Kovalev,[36] and the character table is given in Table EPAPS1.[37] The resulting basis vectors for the two independent cobalt atoms are given in Table 2. The allowed models correspond to an easy-axis ferromagnet ($\Gamma_3^1$), an easy-plane ferromagnet ($\Gamma_9^2$), and two checkerboard antiferromagnets with moments along the $c$-direction ($\Gamma_6^1$) or in the basal plane ($\Gamma_{10}^2$), respectively. The solutions with moments in the basal plane are no longer tetragonal and therefore independent $m_x$, $m_y$ components are allowed.[38] The absence of a measurable lattice distortion, however, prevents their unique determination. In the AF state the decomposition is $\Gamma_{Mag} = 1\Gamma_2^1 + 1\Gamma_7^1 + 1\Gamma_9^2 + 1\Gamma_{10}^2$ and $\Gamma_{Mag} = 1\Gamma_2^1 + 1\Gamma_3^1 + 1\Gamma_9^2 + 1\Gamma_{10}^2$ for the Co and Nd sites, respectively. In case of a 2$^{nd}$ order phase transition, Landau theory states that only a single IR becomes critical. This leaves the $\Gamma_2^1$, $\Gamma_9^2$ and $\Gamma_{10}^2$ symmetries for combined Nd and Co ordering. The unit cell contains two independent Co and two independent Nd sites with basis vectors as given in Table 2. The $\Gamma_2^1$ and $\Gamma_{10}^2$ solutions have FM Co and FM Nd planes (coupled AF) with moments constrained along the $c$-direction and in the basal plane, respectively. The $\Gamma_9^2$ solution has AF Co planes and FM Nd planes (coupled AF). The presence of only magnetic (00l) reflections in the FM state reveals that the ordered moment is constrained to the basal plane, and that the $\Gamma_9^2$ solution is the correct one. Rietveld fitting gives an ordered moment of 0.30(1) $\mu_B$ at 15 K and 0.26(1) $\mu_B$ at 40 K. A representation of the magnetic structure is shown in Fig 1d. Attempts to fit the magnetic intensities to any of the other three models in Table 2 were not successful. Subtraction of the 15 K and 1.7 K data showed that the FM contribution to the (00l) reflections has disappeared at base temperature. Trial Rietveld refinements quickly established that the $\Gamma_2^1$ and $\Gamma_9^2$ models do not give good fits to the observed magnetic intensities. The $\Gamma_{10}^2$ solution in contrast yields an excellent fit to the data. In this model, adjacent FM Nd layers couple AF, and the FM Co layer at z = ¼ couples AF with the FM Co layer at z = ¾. All moments are constrained to the basal plane and Rietveld fitting gives $m_{Co}$ = 0.26(6) $\mu_B$ and $m_{Nd}$ = 1.39(4) $\mu_B$ at 1.7 K. The final Rietveld fit and a schematic representation of the magnetic structure are given in Fig. 1c.

### Structure and properties of NdFe$_{1-x}$Co$_x$AsO (0 < x < 1)

Inspection of the synchrotron X-ray powder diffraction patterns revealed that all prepared samples have the tetragonal P4/nmm structure at RT. The lattice constants were obtained from Rietveld fits, and revealed that the crystallographic c/a-ratio varies linearly with x over the entire composition range (0 ≤ x ≤ 1), as shown in Fig. 2. The solid circles are selected samples where the composition has been confirmed from combined Rietveld fits to synchrotron X-ray and neutron powder diffraction data. The linear dependence suggests that the c/a-ratio can be used as an experimental measure of the composition. For most compositions, the agreement between the nominal and calculated (c/a-derived) compositions is within 1-3 esd's (Table 3). For some samples (e.g. $x_{nominal}$ = 0.10) a larger deviation is found, which signals the formation of impurities that change the stoichiometry of the main phase. Use of the calculated compositions results in a smoother x-dependence of the refined lattice constants, bond lengths and bond angles. In addition, it discriminates between samples with the same nominal composition but different SC transition temperatures, and vice versa (Table 3). For these reasons, the compositions (x) used in this manuscript are the ones calculated from the c/a-ratio unless stated otherwise. The x-dependence of the lattice constants are shown in Fig. 2. The *a*-axis is almost constant at low doping levels, in agreement with other cobalt doped superconductors,[17, 39] and increases moderately for x > 0.25. In contrast, the *c*-axis contracts rapidly. The (Fe/Co)-As bond lengths show a gradual contraction from 2.40 Å for x = 0 to 2.35 Å for x = 1, while the tetrahedral angle increases from 111.4° (x = 0) to 115.9° (x = 1, inset to Fig. 2b).

In order to independently confirm the composition of selected samples, combined Rietveld analysis of room temperature synchrotron X-ray and neutron powder diffraction data was undertaken. This allows the simultaneous refinement of the total transition metal site occupancy and the iron to cobalt ratio. The total occupancy is effectively obtained from fitting the X-ray data as there is no significant scattering contrast between Fe and Co, while the difference in neutron scattering length [Fe: 9.45 fm, Co: 2.49 fm] allows for the determination of the Co/Fe ratio. Four compositions were studied: $x_{nominal}$ = 0.075, 0.125, 0.175 and 1, while the x = 0 composition has been previously reported.[29] In these refinements, the lattice constants were kept at the synchrotron X-ray values and the neutron wavelength (λ = 1.3010(2) Å) was refined to fit the neutron diffraction data. The fit to the $x_{nominal}$ = 0.125 X-ray and neutron powder diffraction data is shown in Fig. 3. The other fits are given in Fig. EPAPS1. The results of the combined Rietveld fits are summarized in Table 1. They reveal that the total occupancy of the transition metal site is unity in all cases, and that the refined Fe and Co fractions are within a single estimated standard deviation of the nominal values. This result is robust as trials with randomly picked Fe and Co fractions always recovered the result shown in Table 3. Refinement of the other site occupancies indicate slightly larger than full occupancies on the As-site and slightly reduced O-site occupancies (Table 1). It is not clear if these are significant but we note that in a simple ionic model with $As^{3-}$ and $O^{2-}$ the charge doping effects cancel out. In addition, oxygen deficient 1111 type samples, such as $NdFeAsO_{1-d}$, have only been prepared using high-pressure high-temperature synthesis routes, suggesting that ambient pressure routes do not normally lead to oxygen deficiencies.[40]

Variable temperature synchrotron X-ray diffraction was used to follow the P4/nmm (T) to Cmma (O) transition associated with the AF ordering that occurs in 1111-type parent materials. In case of NdFeAsO (x = 0), a broad transition with initial broadening of Bragg reflections at ~160 K, followed by a full splitting at ~140 K has been reported.[20, 41] The latter temperature corresponds to the onset of long range AF order of the Fe-spins.[42] For NdFe$_{1-x}$Co$_x$AsO [x = 0.047(2), x = 0.065(2) and x = 0.123(2)], data were collected on heating from 4 K. At low temperatures a broadening of reflections with Miller indexes $h \neq 0$ and $k \neq 0$, consistent with the well established T→O transition, was evident (Fig. EPAPS2). However, no peak splitting occurs down to the lowest measured temperatures. Structural refinements using the orthorhombic Cmma model showed a clear divergence of the *a*- and *b*-axis for x = 0.047(2) and x = 0.065(2), respectively. In contrast, refinements for x = 0.122(2) showed no evidence for a structural distortion as Rietveld fits with the P4/nmm and Cmma structural models gave the same residuals and almost identical *a*- and *b*-axes. The slight broadening evident for x = 0.122(2) in Fig. EPAPS2 can therefore not be attributed to the T→O transition. The temperature dependence of the lattice constants is shown in Fig. 4, where the orthorhombic cell constants ($a_O, b_O$) have been divided by √2. The lattice constants above the T→O transition were obtained from fits to the P4/nmm structural model. The transition temperatures were taken from the onset temperature of orthorhombic strain, s = ($a_O$-$b_O$)/($a_O$+$b_O$), and are 72(2) K [x = 0.047(2)] and 64(2) K [x = 0.065(2)], respectively (Fig. 4). The low field ZFC susceptibilities for the NdFe$_{1-x}$Co$_x$AsO solid solution series are given in Fig. 5. Superconducting transitions are observed for samples with 0.065(2) ≤ x ≤ 0.172(2). The transition temperatures were taken from the diamagnetic onset temperature and are summarized in Table 3. This reveals almost identical values of 16.5(2) K for optimally doped samples near x = 0.12 [x = 0.118(2), x = 0.122(2) and x = 0.123(2)]. The diamagnetic shielding fractions vary from 30-70% of 4πχ for samples with x = 0.12 and are lower on the under and overdoped sides as are the critical temperatures. The SC transitions were confirmed by electrical resistance measurements. The normalized resistances $R/R_{300K}$ are given in Fig. 6. The 300 K resistivities fall between 5 mΩ cm for x = 0 and 2 mΩ cm for x = 1. The normal state resistance for x = 0 shows the familiar drop below T ~ 160 K with the maximum slope dR/dT at 140 K. These temperatures correspond to the onset of the T→O transition and SDW order, respectively.[20] Samples with x > 0.2 have metallic temperature dependences, indicating a transition to more conventional metallic behavior on the overdoped side. On the underdoped side, increasing x suppresses the SDW transition and no anomaly in R(T) is evident for any of the SC samples. At low temperatures SC transitions are evident. The transition temperatures were taken from the 50% of normal state resistance values (0.5 $R_N$) and are listed in Table 3. The obtained values are in good agreement with the values from the magnetic susceptibility measurements. The widths of the resistive transitions were determined by taking the difference between 0.9 $R_N$ and 0.1 $R_N$, and are ~2 K wide for the optimally doped samples (x = 0.12), ~5 K for x = 0.065(2), and ~3 K for x = 0.0172(2). This is comparable to other polycrystalline 1111-type cobalt doped superconductors including $LaFe_{1-x}Co_xAsO$ (Ref. 13, 17) and $SmFe_{1-x}Co_xAsO$ (Ref. 17), and polycrystalline samples of $Ba(Fe_{1-x}Co_x)_2As_2$.[9] The field dependence of R(T) in the vicinity of the SC transition was studied up to 9 Tesla for samples with x = 0.065(2), x = 0.118(2) and x = 0.172(2). The results are shown in the left panels of Fig. 7, and reveal a significant suppression of $T_c$ with applied magnetic field. The upper critical fields were determined at the temperatures of the 10%, 50% and 90% reduction of the normal state resistance. The results are plotted in the right hand side panels of Fig. 7. In all samples, an initial rapid reduction in small applied fields $\mu_0H < 0.5$ Tesla is evident. This is followed by an approximately linear decrease for $\mu_0H > 1$ Tesla, which shows no sign of saturation. The slopes are dependent on the used resistance criterion, signaling a broadening of the transition in applied magnetic fields. This is characteristic for type-II superconductors, and was also observed in the $LaFe_{1-x}Co_xAsO$ series.[13] Using the 0.5 $R_N$ criterion, almost identical slopes of -2.5(1) T K$^{-1}$ [x = 0.065(2)], -2.3(1) T K$^{-1}$ [x = 0.118(2)] and - 2.3(1) T K$^{-1}$ [x = 0.172(2)] were obtained from linear fits between $1 \leq \mu_0 H \leq 9$ T. The linear dependence of H$_{c2}$(T) near T$_c$ suggests that the Werthamer-Helfand-Hohenberg (WHH) model, which predicts H$_{c2}$(0) = 0.69 T$_c$[dH$_{c2}$/dT], can be used to estimate the upper critical field at zero temperature.[43] This yields the following estimates: H$_{c2}$(0) = 14(1) T for x = 0.065(2), H$_{c2}$(0) = 26(1) T for x = 0.118(2) and H$_{c2}$(0) = 21(1) T for x = 0.172(2). The SC coherence length ($\xi$) is given by $\xi^2(0) = \Phi_0/2\pi H_{c2}(0)$, where $\Phi_0 = 2.07 \times 10^{-7}$ Oe cm$^2$, yielding a coherence length of 36(1) Å for x = 0.118(2).

## Discussion

Like the isostructural RFeAsO high-T$_c$ parent materials NdCoAsO exhibits significant interplay between the rare-earth and transition metal sublattices. Below 85 K, NdCoAsO is an itinerant FM with a small ordered cobalt moment of 0.3 $\mu_B$ constrained to the *ab*-plane. This is analogous to LaCoAsO, which has a Curie temperature of ~60 K and a moment of 0.5 $\mu_B$,[13, 22] and consistent with calculations for LaCoAsO that reveal a spin-polarized band structure with a small ordered moment.[22] Our Arrott plots suggest a linear dependence of M$^2$ on H/M in large fields, typical of weak itinerant ferromagnets such as ZrZn$_2$ but different to the "anomalous" M$^4$ dependence reported for LaCoAsO.[23] More extensive measurements are needed to investigate this. Below 9 K, Nd spin-ordering transforms NdCoAsO to an AF metal. Both the FM and AF transitions are evident in the temperature derivative of the resistance but do not result in large anomalies as observed for RFeAsO. In the RFeAsO materials the Fe spins order in a striped pattern with a small ordered moment of ~ 0.4 $\mu_B$ for all R from spectroscopic methods such as Mossbauer and μSR.[25] Neutron powder diffraction, in contrast, suggests a wider spread of 0.3-0.8 $\mu_B$,[26] which Maeter et al. have attributed to an induced magnetization on the R sublattice below T$_{SDW}$ that contributes to the "Fe" magnetic Bragg reflections.[25] It is not clear from our measurements whether this is important but the almost constant Co moment between 1.7 and 40 K (~ 0.3 $\mu_B$) suggests that there is no significant induced "Nd-contribution". The spectroscopic measurements also suggest that the R spin-ordering does not have any effect on the magnitude of the Fe moment, while for neutron powder diffraction increases have been reported, in particular for R = Nd.[42, 44] In the current case, the available evidence points towards an unchanged Co moment. However, we note that in symmetry unrestricted refinements (i.e. by allowing canting out of the basal plane) it is possible to obtain solutions with $m_{Co}$ = 0.9(2) $\mu_B$ but these are characterized by large correlations and large estimated standard deviations. Trial refinements with non-collinear R ordering analogous to that reported for CeFeAsO (Ref. 8) did not give satisfactory fits to the data. The fitted Nd moment is 1.39(4) $\mu_B$, which is comparable to the value for NdFeAsO (1.55(4) $\mu_B$).[29, 44]

The lattice constants, bond distances and angles for $NdFe_{1-x}Co_xAsO$ change gradually, signaling the formation of a solid solution, as confirmed by Rietveld refinement of the Fe/Co occupancies of selected samples against neutron and synchrotron X-ray data. Furthermore, the structural analysis reveals that the c/a-ratio can be used as an experimental measure of the composition (Fig. 2a). This is valid because the combined X-ray and neutron refinements presented in Table 1 confirm that the only significant compositional change throughout the series is the Fe/Co ratio. The two-fold As-Fe-As tetrahedral angle ($\alpha$, defined in Fig. 9) increases with x. This is in contrast to most other forms of chemical doping, and may in part explain why cobalt doping has not yielded the high $T_c$'s observed in other electron doped superconductors such as $NdFeAsO_{1-x}F_x$ and $NdFeAsO_{1-d}$, despite having the same nominal amount of doping. The chemical disorder introduced by mixing Fe/Co on the same crystallographic site is also expected to have a detrimental effect on the maximum attainable $T_c$'s. The variable temperature synchrotron diffraction study reveals subtle structural distortions consistent with the well established T→O structural transition upon cooling for samples with x = 0.046(2) and x = 0.065(2). This confirms that replacement of Fe with Co suppresses the T→O transition and the associated SDW.

The phase diagram for low cobalt substitutions is presented in Fig. 8, and reveals the presence of a SC dome with limiting compositions $0.05 < x < 0.20$. The maximum critical temperature is 16.5(2) K for $x = 0.12$. The underdoped samples ($x \leq 0.075$) have the orthorhombic Cmma structure. The presence of a "dome" and possible overlap between magnetic and SC regions is consistent with results published for other 1111- and 122-type cobalt doped superconductors although the existence of phase coexistence is still under debate.[9, 17, 18, 45] The critical field, $H_{c2}(T)$, increases by approximately 2.3(1) T K$^{-1}$ for the three measured compositions upon cooling below $T_c$. This is smaller than for "indirect" (not Fe-site) electron doped superconductors, such as NdFeAsO$_{1-x}$F$_x$, where values of 4-5 T K$^{-1}$ have been observed.[46] However, similar values are reported for other cobalt doped superconductors,[13, 39] suggesting that the superconductivity in these materials is less robust against applied magnetic fields.

Finally, the maximum observed critical temperatures ($T_{c, max}$) for well characterized cobalt doped superconductors are plotted against the tetrahedral As-Fe-As angle ($\alpha$). The angles used are for the parent materials as $\alpha$ is not commonly reported for doped compositions. However, this does not affect the reported trends significantly as the increase in $\alpha$ upon cobalt doping is expected to be similar in all materials. The following observations can be made: (1) $T_{c,max}$ increases as $\alpha$ tends towards the cubic value. This is in agreement with the literature for indirectly doped superconductors,[8, 19, 47] although there is no evidence for a maximum occurring at the ideal tetrahedral angle. (2) The 122-type materials have higher $T_c$'s than the 1111-type for a given value of $\alpha$. This contrasts the situation for indirect doping where the largest $T_c$'s up to 55 K are obtained for electron doped 1111-phases such as NdFeAsO$_{1-x}$F$_x$, while the hole doped 122 phases with a smaller separation between Fe$_2$As$_2$ planes are limited to 38 K for doped Ba$_{1-x}$K$_x$Fe$_2$As$_2$ (both parents have $\alpha \sim 111.2°$). This suggests that that were it possible to electron dope the 122 superconductors via chemical substitutions on the AE-site higher Tc's than 55 K may be achieved.

To summarize, we have investigated the structures and properties of the NdFe$_{1-x}$Co$_x$AsO series ($0 \leq x \leq 1$). For low cobalt doping, the phase diagram contains a superconducting dome ($0.05 < x < 0.20$, maximum $T_c = 16.5(2)$ K for $x = 0.12$). Samples with $x \leq 0.075$ have the orthorhombic Cmma structure at low temperatures. NdCoAsO is an itinerant ferromagnet ($T_C = 85$ K) with a small cobalt moment (0.3 $\mu_B$) that shows a transition to an antiferromagnetic state at $T_N = 9$ K.

**Acknowledgements**

JWGB acknowledges the Royal Society of Edinburgh for financial support and the EPSRC for the provision of beam time at the ILL and ESRF.

Table 1. Lattice constants, refined atomic parameters, selected bond distances and angles, and fit statistics for the combined Rietveld fits to synchrotron X-ray and neutron powder diffraction data for selected NdFe$_{1-x}$Co$_x$AsO compositions.

| | $x_{nominal}$ | 0.075 | 0.125 | 0.175 | 1.00 |
|---|---|---|---|---|---|
| | $a$-axis (Å) | 3.96754(1) | 3.96711(3) | 3.96759(4) | 3.98724(1) |
| | $c$-axis (Å) | 8.58069(5) | 8.5611(1) | 8.5462(1) | 8.31835(4) |
| | Volume (Å$^3$) | 135.072(1) | 134.734(1) | 134.532(1) | 132.246(1) |
| Nd | $U_{iso}$ (Å$^2$) | 0.0041(1) | 0.0046(1) | 0.0049(1) | 0.0006(7) |
| | z | 0.13911(3) | 0.13962(4) | 0.14004(6) | 0.14221(5) |
| | Occ. | 1.00 | 1.00 | 1.00 | 1.00 |
| Fe/Co | $U_{iso}$ (Å$^2$) | 0.0049(1) | 0.0050(2) | 0.0049(3) | 0.0066(2) |
| | Fe - Occ. | 0.918(4) | 0.886(6) | 0.827(8) | 0 |
| | Co - Occ. | 0.076(4) | 0.123(6) | 0.180(8) | 0.996(2) |
| | Tot - Occ. | 0.994(5) | 1.009(8) | 1.007(11) | 0.996(2) |
| As | $U_{iso}$ (Å$^2$) | 0.0057(1) | 0.0057(1) | 0.0060(2) | 0.0059(1) |
| | z | 0.65681(5) | 0.65577(8) | 0.6549(1) | 0.65066(8) |
| | Occ. | 1.011(2) | 1.024(3) | 1.019(3) | 1.003(2) |
| O | $U_{iso}$ (Å$^2$) | 0.0047(3) | 0.0048(4) | 0.0029(6) | 0.0036(3) |
| | Occ. | 0.978(4) | 0.962(6) | 0.980(8) | 1.000(4) |
| | $x_{Rietveld}$ | 0.076(4) | 0.122(6) | 0.178(8) | 0.996(2) |
| | d(Fe/Co-As) (Å) | 2.3970(3) | 2.3902(4) | 2.3850(5) | 2.3546(3) |
| | ∠As-(Fe/Co)-As (°) | 111.70(2) | 112.17(3) | 112.56(4) | 115.70(3) |
| | ∠As-(Fe/Co)-As (°) | 108.37(1) | 108.14(1) | 107.95(2) | 106.45(1) |
| | $\chi^2$ | 3.1 | 3.2 | 5.2 | 3.1 |
| NPD | $wR_p$ | 3.6 | 3.6 | 3.3 | 3.0 |
| | $R_p$ | 2.5 | 2.5 | 2.4 | 2.3 |
| | $R_F^2$ | 2.9 | 2.5 | 2.8 | 3.0 |
| XRD | $wR_p$ | 7.9 | 8.7 | 12.9 | 10.1 |
| | $R_p$ | 5.4 | 6.2 | 8.7 | 6.8 |
| | $R_F^2$ | 2.5 | 2.7 | 7.8 | 3.5 |
| | NdAs (wt%) | 1.4(1) | 6.3(1) | 3.4(1) | 8.6(1) |
| | CoAs (wt%) | - | - | - | 7.3(3) |
| | Nd$_2$O$_3$ (wt%) | 1.1(1) | 3.4(1) | 2.3(1) | - |

Space group P4/nmm, Nd on 2$c$(¼, ¼, z); Fe/Co on 2$b$(¾, ¼, ½); As on 2$c$(¼, ¼, z); O on 2$a$(¼, ¾, 0).

Table 2. Basis vectors [$m_x$, $m_y$, $m_z$] for the space group P4/nmm with **k** = 0 and **k** = (0 0 ½).

| | **k** = 0 | | | | **k** = (0 0 ½) | | |
|---|---|---|---|---|---|---|---|
| | $\Gamma_3^1$ | $\Gamma_6^1$ | $\Gamma_9^2$ | $\Gamma_{10}^2$ | $\Gamma_2^1$ | $\Gamma_9^2$ | $\Gamma_{10}^2$ |
| Co1 | [0 0 $m_z$] | [0 0 $m_z$] | [$m_x$ $m_y$ 0] | [$m_x$ $m_y$ 0] | [0 0 $m_z$] | [$m_x$ $m_y$ 0] | [$m_x$ $m_y$ 0] |
| Co2 | [0 0 $m_z$] | [0 0 -$m_z$] | [$m_x$ $m_y$ 0] | [-$m_x$ -$m_y$ 0] | [0 0 $m_z$] | [-$m_x$ -$m_y$ 0] | [$m_x$ $m_y$ 0] |
| Nd1 | - | - | - | - | [0 0 $m_z$] | [$m_x$ $m_y$ 0] | [$m_x$ $m_y$ 0] |
| Nd2 | - | - | - | - | [0 0 $m_z$] | [-$m_x$ -$m_y$ 0] | [$m_x$ $m_y$ 0] |

Co1: (¾, ¼, ½), Co2: (¾, ¼, ½), Nd1: (¼, ¼, 0.14), Nd2: (¼, ¼, 0.86), coordinates in crystallographic cell.

Table 3. Crystallographic c/a-ratios, compositions from the c/a-ratio (x), compositions from Rietveld fits ($x_{Rietveld}$), and critical temperatures from the onset of diamagnetism and resistive transition midpoint for the NdFe$_{1-x}$Co$_x$AsO series.

| $x_{nominal}$ | c/a | $x^{*1}$ | $x_{Rietveld}$ | $T_c, \chi_{onset}$ (K) | $T_c, R_{mid}$ (K) |
|---|---|---|---|---|---|
| 0 | 2.16799(4) | 0 | | | |
| 0.05 | 2.16420(4) | 0.047(2) | | <1.8 | <1.8 |
| 0.075 (NPD) | 2.16272(4) | 0.065(2) | 0.076(4) | 12.8(2) | 9.2(2) |
| 0.075 | 2.16180(4) | 0.076(2) | | 15.0(2) | |
| 0.10 | 2.15808(4) | 0.122(2) | | 16.7(2) | 16.2(2) |
| 0.125 | 2.15840(4) | 0.118(2) | | 16.5(2) | 16.0(2) |
| 0.125 (NPD) | 2.15802(4) | 0.123(2) | 0.122(6) | 16.6(2) | |
| 0.15 | 2.15546(4) | 0.154(2) | | 15.1(2) | |
| 0.175 (NPD) | 2.15398(4) | 0.172(2) | 0.178(8) | 13.4(2) | 12.6(2) |
| 0.2 | 2.15121(4) | 0.206(2) | | <1.8 | <1.8 |
| 0.25 | 2.14726(4) | 0.255(2) | | <1.8 | <1.8 |
| 0.50 | 2.12845(4) | 0.486(2) | | | |
| 1 | 2.08657(4) | 1 | 0.996(2) | | |

*1 c/a = -0.0814(5)x + 2.16799(4)

**Figure captions**

Fig. 1. This figure summarizes the measurements on NdCoAsO. Panel (**a**) shows the temperature dependence of the ZFC magnetic susceptibility ($\chi$) and electrical resistivity ($\rho$), and their temperature derivatives. Panel (**b**) shows Arrott plots derived from the isothermal M(H) measurements (shown in the inset). Panel (**c**) shows the Rietveld fits to the 1.7 K neutron powder diffraction data (Fit statistics: $wR_p$ = 3.5%, $R_p$ =2.3%, $R_F^2$ = 8.4%) and a representation of the magnetic structure at that temperature. The tick marks are from top to bottom: magnetic phase, NdCoAsO, CoAs and NdAs. Panel (**d**) shows the Rietveld fit to the 15 K -100 K difference neutron powder diffraction pattern and a schematic representation of the fitted ferromagnetic magnetic structure.

Fig. 2 (**a**) Left axis: crystallographic c/a-ratio versus $x_{nominal}$ for NdFe$_{1-x}$Co$_x$AsO. The fitted line is c/a = -0.0814(5)x + 2.16799(4). Right axis: the *a*-axis versus the composition derived from the c/a-ratio (x). A fit to (**b**) Crystallographic *c*-axis plotted versus x. The insets show the x-dependence of the Fe-As bond distance and the tetrahedral As-Fe-As angle ($\alpha$, defined in the inset of Fig. 9) for the NdFe$_{1-x}$Co$_x$AsO series. The solid lines are guides to the eye.

Fig. 3. Combined Rietveld fit to room temperature synchrotron X-ray (**a**) and neutron powder diffraction data (**b**) for a sample with $x_{nominal}$ = 0.125. Observed data are indicated by open circles, the fit by the red line and the difference curve by the green line. Bragg positions are indicated by vertical markers. The bottom markers correspond to NdAs and Nd$_2$O$_3$ impurities (Table 1).

Fig. 4. Temperature dependence of the crystallographic basal plane lattice constants for selected NdFe$_{1-x}$Co$_x$AsO compositions.

Fig. 5. Temperature dependence of the low field magnetic susceptibilities for NdFe$_{1-x}$Co$_x$AsO. The inset shows x-dependence of the diamagnetic fraction at 2 K.

Fig. 6. Temperature dependence of the normalized electrical resistance (R/R$_{300K}$) for NdFe$_{1-x}$Co$_x$AsO. The inset illustrates the superconducting transitions.

Fig. 7. Temperature and field dependence of the electrical resistance for selected NdFe$_{1-x}$Co$_x$AsO compositions. H$_{c2}$(T) derived from 10% (0.1 R$_N$), 50% (0.5 R$_N$) and 90% (0.9 R$_N$) resistance drops are shown on the right hand side.

Fig. 8. Temperature-composition phase diagram for NdFe$_{1-x}$Co$_x$AsO showing the suppression of the structural T→O phase transition with increasing Co concentration and the superconducting dome. The data points for the T→O structural transition (T$_s$) were obtained from synchrotron X-ray powder diffraction, while the superconducting T$_c$ values were obtained from the diamagnetic onset temperatures (T$_\alpha$, red circles) and midpoint resistances (T$_\beta$, black squares). Error bars indicate 10% and 90% of the resistive transition. The green dotted line is a linear fit to the T$_s$ data.

Fig. 9. Variation of the superconducting T$_c$ with tetrahedral As-Fe-As angle (defined in inset) for well characterized 122-type AE(Fe$_{1-x}$Co$_x$)$_2$As$_2$ (solid circles), and 1111-type RFe$_{1-x}$Co$_x$AsO and AEFe$_{1-x}$Co$_x$AsF (open squares) superconductors. The line is intended as a guide to the eye. Data were taken from the following references: Ca[F],[18] La[O],[13, 17] Sm[O],[17] Ca,[15] Sr,[16] and Ba.[39]

Fig. 1.

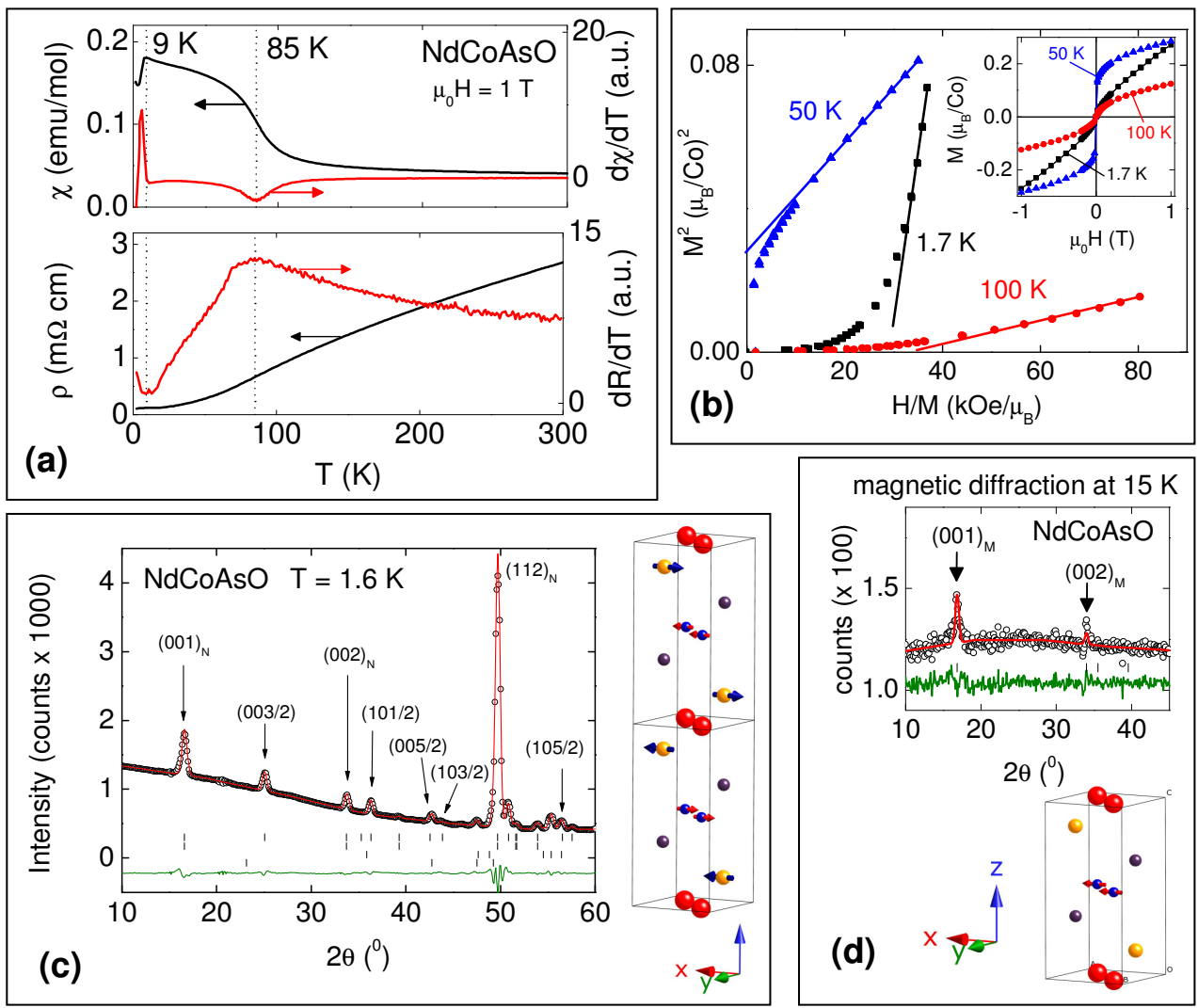

Fig. 2.

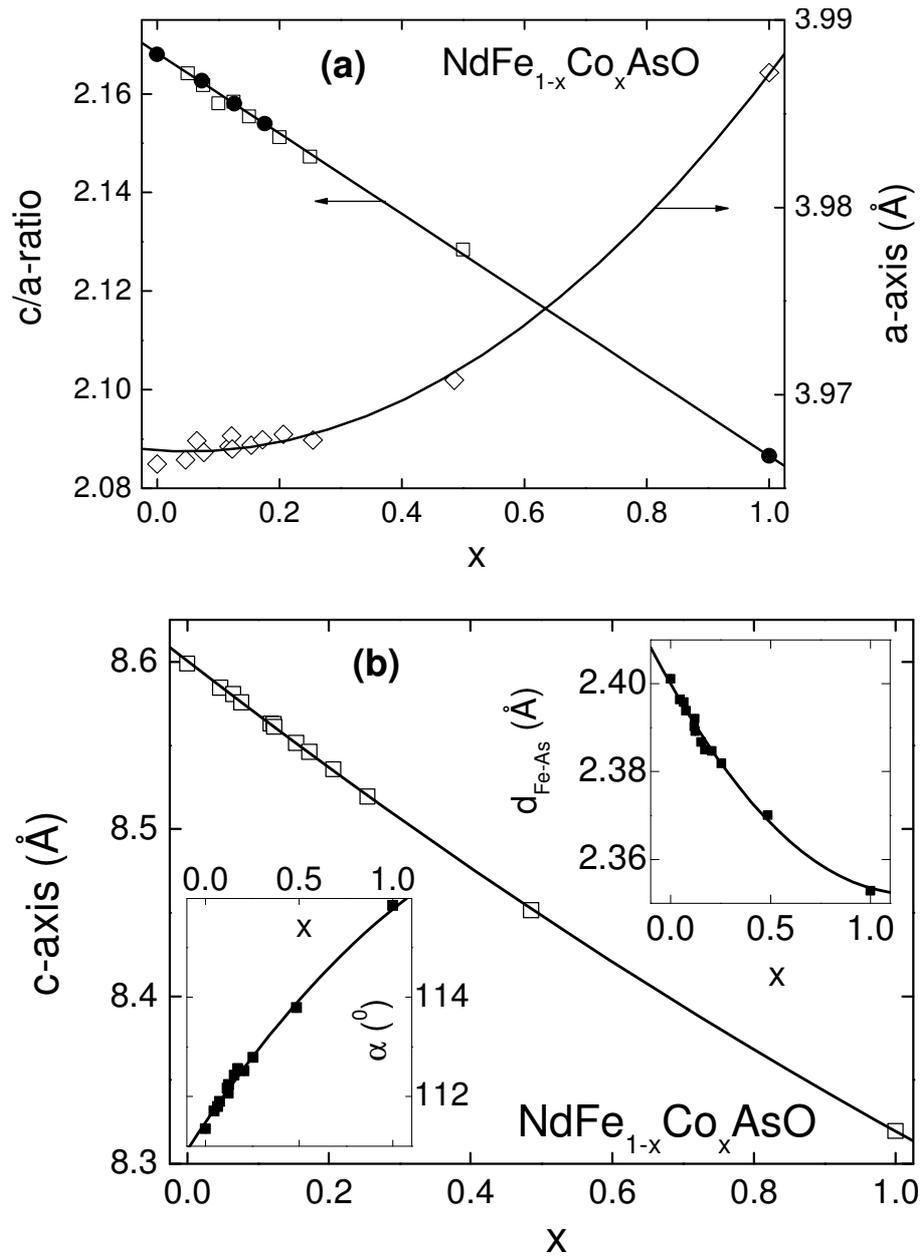

Fig. 3.

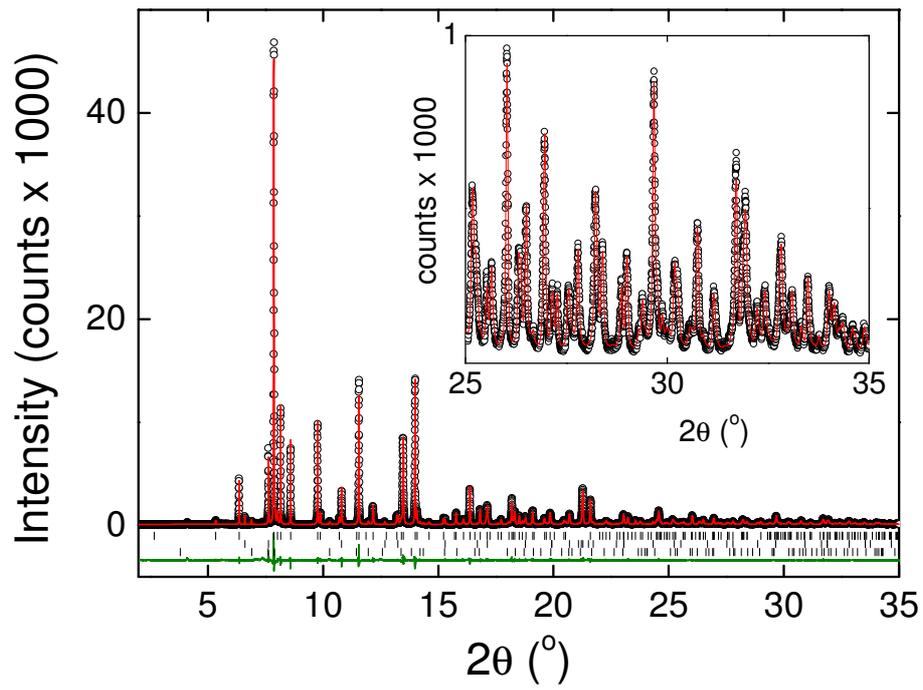

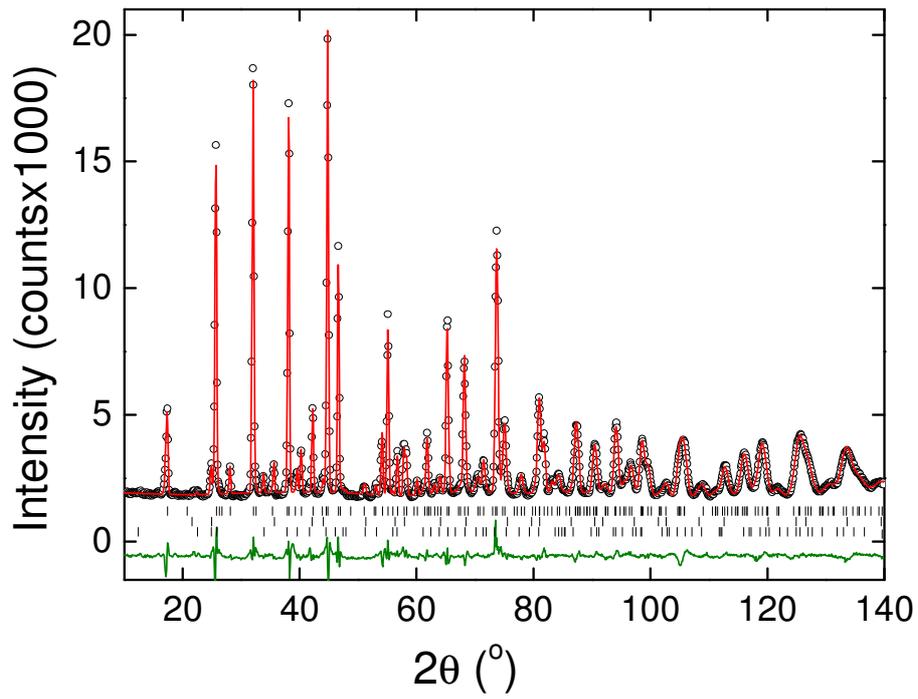

Fig. 4.

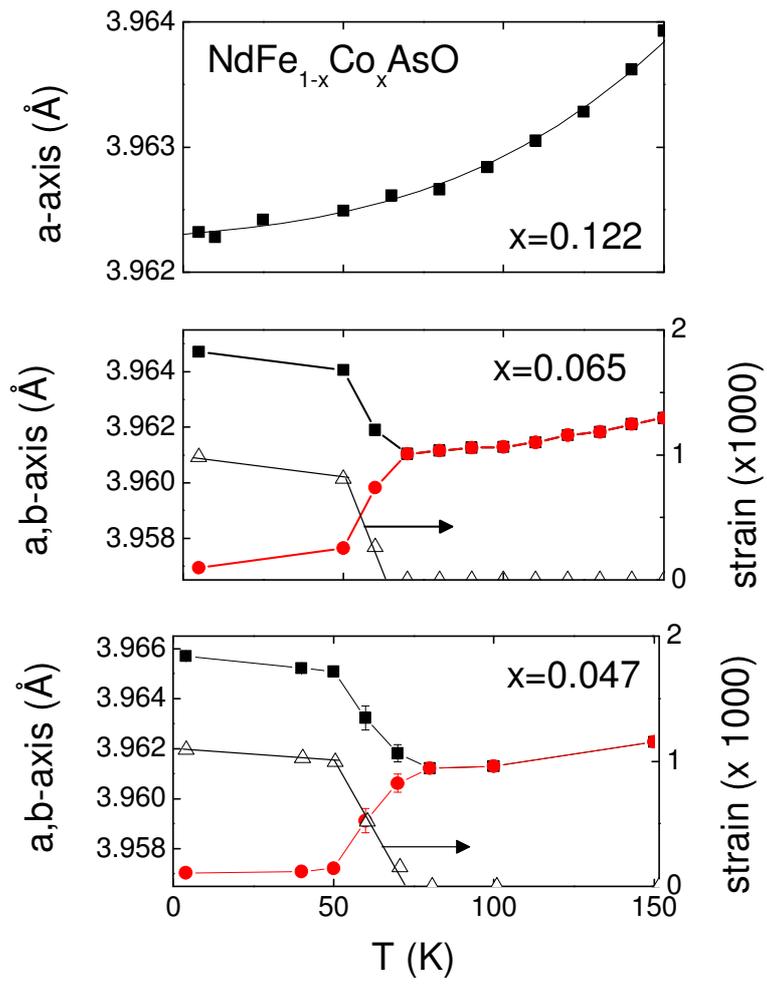

Fig. 5

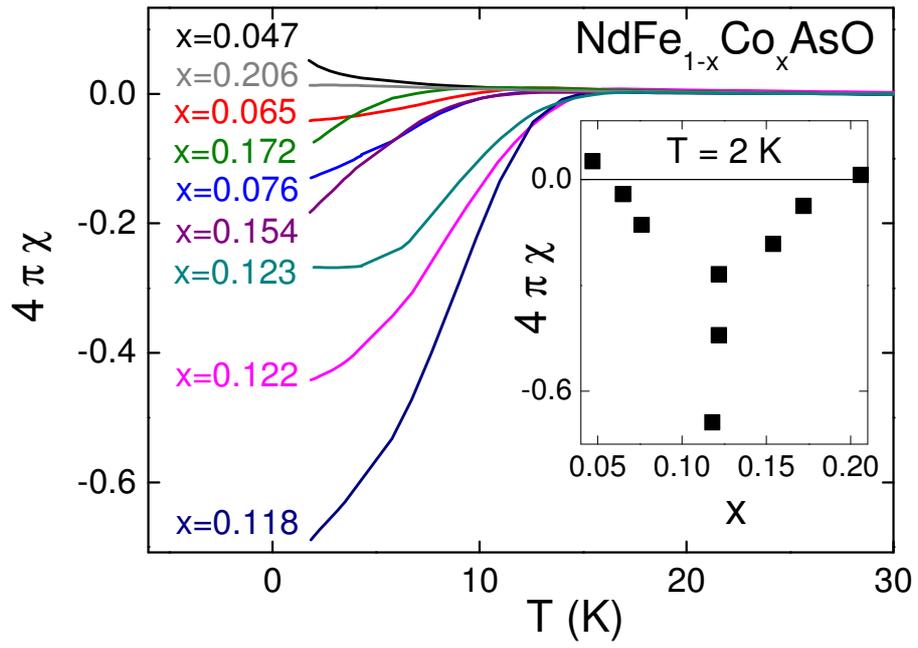

Fig. 6.

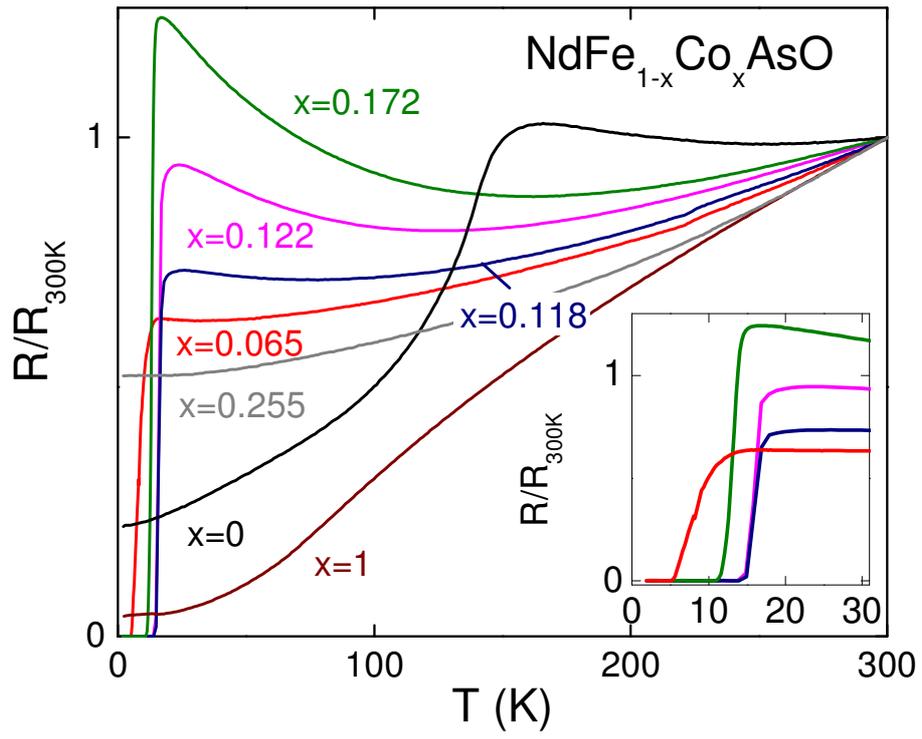

Fig. 7.

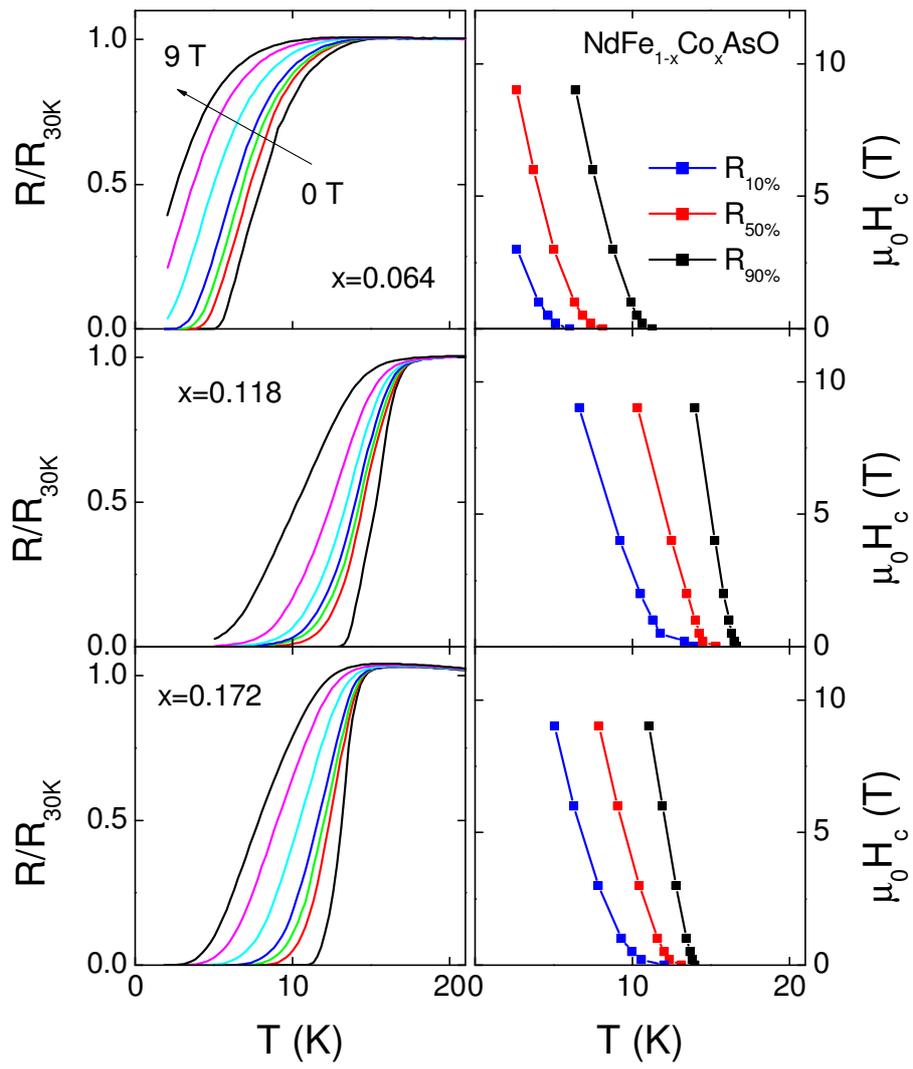

Fig. 8.

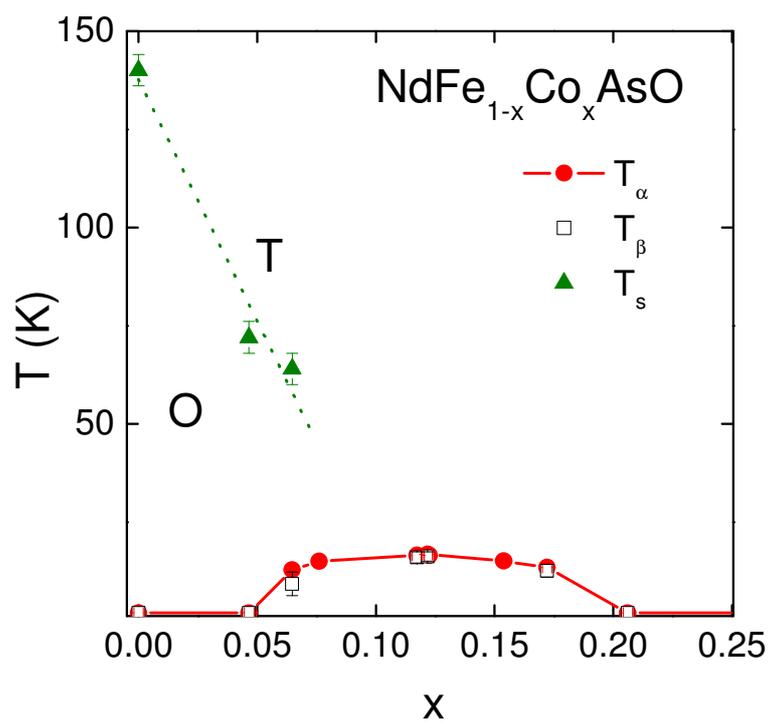

Fig. 9

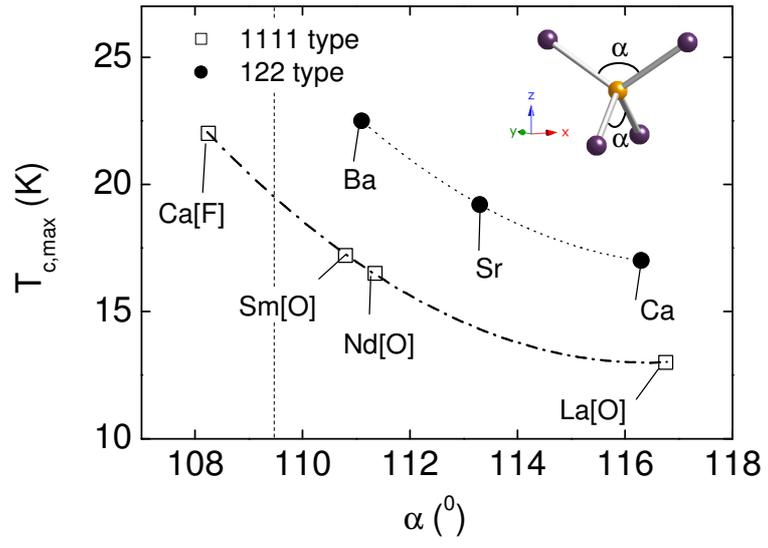

EPAPS for

# Superconductivity in NdFe$_{1-x}$Co$_x$AsO (0.05 < x < 0.20) and rare-earth magnetic ordering in NdCoAsO

Andrea Marcinkova[1], David A. M. Grist[1], Irene Margiolaki[2], Thomas C. Hansen[3], Serena Margadonna[1] and Jan-Willem G. Bos[1,4]

1. School of Chemistry and Centre for Science at Extreme Conditions, University of Edinburgh, United Kingdom, EH9 3JJ.
2. European Synchrotron Radiation Facility, 39043 Grenoble, France.
3. Institute Laue Langevin, 38042 Grenoble, France.
4. Department of Chemistry – EPS, Heriot-Watt University, Edinburgh, United Kingdom, EH14 4AS

Table EPAPS1. Character table for small group $G_k$ and space group P4/nmm.

| | E | 2x | 2y | 2z | 2b | 4z$^+$ | 4z$^-$ | 2a | I | σ$_x$ | σ$_y$ | σ$_z$ | σ$_{db}$ | S4z$^-$ | S4z$^+$ | σ$_{da}$ |
|---|---|---|---|---|---|---|---|---|---|---|---|---|---|---|---|---|
| $\Gamma_1^1$ | 1 | 1 | 1 | 1 | 1 | 1 | 1 | 1 | 1 | 1 | 1 | 1 | 1 | 1 | 1 | 1 |
| $\Gamma_2^1$ | 1 | 1 | 1 | 1 | 1 | 1 | 1 | 1 | -1 | -1 | -1 | -1 | -1 | -1 | -1 | -1 |
| $\Gamma_3^1$ | 1 | -1 | -1 | 1 | -1 | 1 | 1 | -1 | 1 | -1 | -1 | 1 | -1 | 1 | 1 | -1 |
| $\Gamma_4^1$ | 1 | -1 | -1 | 1 | -1 | 1 | 1 | -1 | -1 | 1 | 1 | -1 | 1 | -1 | -1 | 1 |
| $\Gamma_5^1$ | 1 | 1 | 1 | 1 | -1 | -1 | -1 | -1 | 1 | 1 | 1 | 1 | -1 | -1 | -1 | -1 |
| $\Gamma_6^1$ | 1 | 1 | 1 | 1 | -1 | -1 | -1 | -1 | -1 | -1 | -1 | -1 | 1 | 1 | 1 | 1 |
| $\Gamma_7^1$ | 1 | -1 | -1 | 1 | 1 | -1 | -1 | 1 | 1 | -1 | -1 | 1 | 1 | -1 | -1 | 1 |
| $\Gamma_8^1$ | 1 | -1 | -1 | 1 | 1 | -1 | -1 | 1 | -1 | 1 | 1 | -1 | -1 | 1 | 1 | -1 |
| $\Gamma_9^2$ | 2 | 0 | 0 | -2 | 2 | 0 | 0 | -2 | 2 | 0 | 0 | -2 | 2 | 0 | 0 | -2 |
| $\Gamma_{10}^2$ | 2 | 0 | 0 | -2 | 2 | 0 | 0 | -2 | -2 | 0 | 0 | 2 | -2 | 0 | 0 | 2 |

E (x,y,z), 2x (x,-y,-z), 2y (-x,y,-z), 2z (-x,-y,z), 2b (-y,-x,-z), 4z$^+$ (-y,x,z), 4z$^-$ (y,-x,z), 2a (y,x,-z), I (-x,-y,-z), σ$_x$ (-x,y,z), σ$_y$ (x,-y,z), σ$_z$ (x,y,-z), σ$_{db}$ (y,x,z), S4z$^-$ (y,-x,-z), S4z$^+$ (-y,x,-z), σ$_{da}$ (-y,-x,z).

Fig. EPAPS1. Combined Rietveld fit to room temperature synchrotron X-ray and neutron powder diffraction data for the NdFe$_{1-x}$Co$_x$AsO compositions in Table 1. Observed data are indicated by open circles, the fit by the red line and the difference curve by the green line. Bragg positions are indicated by vertical markers. They correspond (top to bottom) to the main phase, NdAs and Nd$_2$O$_3$ for $x = 0.075$ and $x = 0.175$. For $x = 1.00$, the middle (bottom) row corresponds to CoAs (NdAs). The $x = 0.175$ SXRD pattern contains additional broad reflections marked by asterisks (~5 x the FWHM of the main phase, and ~1/3 of the intensity of the most intense NdAs peak) that remain unaccounted for, and are not evident in the NPD data. Minor traces of this phase are also evident in the other SXRD patterns.

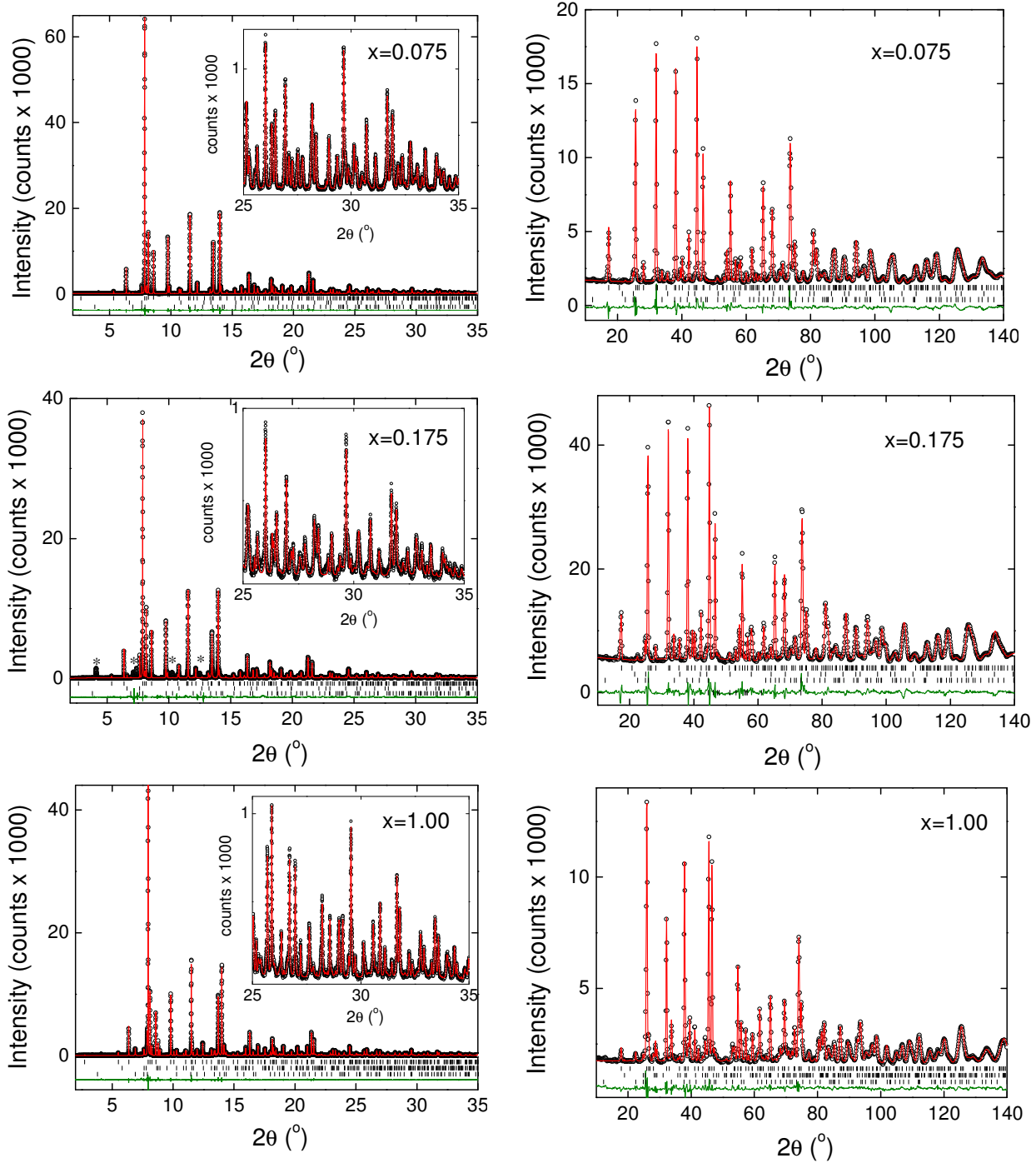

Fig. EPAPS2. Broadening of the tetragonal (110) reflection for NdFe$_{1-x}$Co$_x$AsO indicative of the P4/nmm → Cmma structural transition.

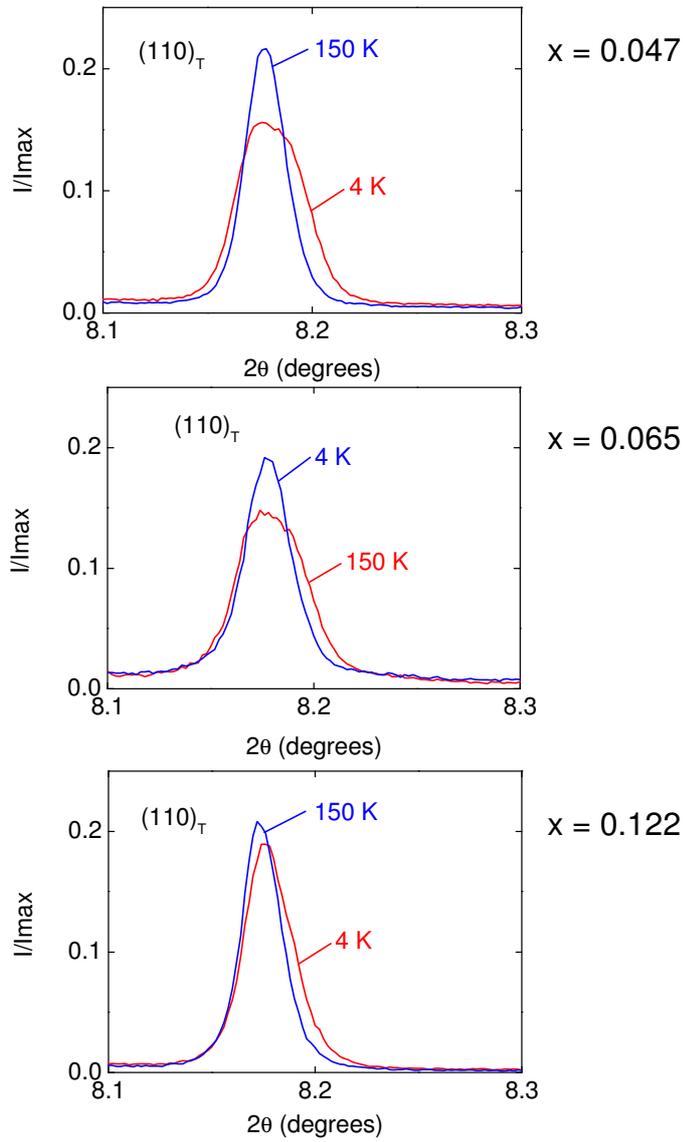